\documentclass[a4paper,11pt]{article}

\usepackage[cp1250]{inputenc}

\newtheorem{dfn}{Definition}[section]
\newtheorem{tw}[dfn]{Theorem}

\newtheorem{rem}[dfn]{Remark}

\newtheorem{lem}[dfn]{Lemma}

\usepackage{anysize}
\usepackage{verbatim}
\usepackage{array}
\usepackage{enumerate}
\usepackage{amssymb} % symbole AMS
\usepackage{amsmath}
\usepackage{fancyhdr}
\usepackage{euscript}
\usepackage{latexsym}

\author{Micha\l \ Barski \\ \small  Faculty of Mathematics and Computer Science, University of Leipzig, Germany\\
\small Faculty of Mathematics, Cardinal Stefan Wyszy\'nski University in Warsaw, Poland
\\ \small{\it Michal.Barski@math.uni-leipzig.de}}

\title{\bf Large losses - probability minimizing approach}
\begin{document}

\maketitle
\renewcommand{\abstractname}{Abstract}
\begin{abstract}
The probability minimizing problem of large losses of portfolio in
discrete and continuous time models is studied. This gives a
generalization of quantile hedging presented in [3].
\end{abstract}
\begin{quote}
\end{quote}
\begin{quote}
\end{quote}
\begin{quote}
{\bf GEL Classification Numbers}: G11, G13
\end{quote}

\section{Introduction}
   Let $(S_t)$ be a $d$-dimensional semimartingale on a filtered
probability space $(\Omega,\mathcal{F},(\mathcal{F}_t),P)$ which
represents the stock prices. We denote by $\mathcal{Q}$ the set of
all martingale measures, it means that $Q\in\mathcal{Q}$ if $Q\sim
P$ and $(S_t)$ is a martingale with respect to $Q$. Let $H$ be a
$\mathcal{F}$ measurable random variable called contingent claim.
It is known that on such market we have two prices: the buyer's
price $u_b=\inf_{Q\in\mathcal{Q}}\mathbf{E}^Q[H]$ and the seller's
price $u_s=\sup_{Q\in\mathcal{Q}}\mathbf{E}^Q[H]$, which usually
are different. A natural question arises : what price from the so
called arbitrage-free interval $[u_b,u_s]$ should be chosen? This
problem was a motivation for introducing risk measures on
financial markets. Various approaches were presented to answer
this question, see for instance [1],[2],[3],[4],[6].

In [3] F\"{o}llmer, and Leukert study the quantile hedging
problem. They define a random variable $\varphi_{x,\pi}$ connected
with the strategy $(x,\pi)$ by:
\begin{gather*}
\varphi_{x,\pi}=\mathbf{1}_{\{X_T^{x,\pi}\geq
H\}}+\frac{X_T^{x,\pi}}{H}\mathbf{1}_{\{X_T^{x,\pi}< H\}},
\end{gather*}
where $X_T^{x,\pi}$ is the terminal value of the portfolio
connected with the strategy $\pi$ starting from the initial
endowment $x$. If $x\geq u_s$ then for the hedging strategy
$\tilde{\pi}$ we have $\mathbf{E}[\varphi_{x,\tilde{\pi}}]=1$,
otherwise $\mathbf{E}[\varphi_{x,\pi}]<1$ for each $\pi$. The aim
of the trader is to maximize $\mathbf{E}[\varphi_{x,\pi}]$ over
$\pi$ from the set of all admissible strategies. Actually, the
motivation of quantile hedging was a slightly different problem,
namely
\begin{gather*}
P(X_T^{x,\pi}\geq H)\longrightarrow \max_{\pi}.
\end{gather*}
This problem was solved by the above approach only in a particular
case.

Now assume that investor has a loss function
$u:[0,\infty)\longrightarrow[0,\infty)$, $u(0)=0$, which is
assumed to be continuous and strictly increasing, and he accepts
small losses of the portfolio. It means he has no objections to
losses s.t. $u((H-X_T^{x,\pi})^+)\leq \alpha$, where $\alpha\geq0$
is a level of acceptable losses fixed by the investor. He wants to
avoid losses which exceed $\alpha$. As the optimality criterion we
admit maximizing probability that losses are small. More
precisely, the problem is
\begin{gather*}
P[u((H-X_T^{x,\pi})^{+})\leq \alpha]\longrightarrow\max_{\pi},
\end{gather*}
where $\pi$ is an admissible strategy. Notice that for $\alpha=0$
we obtain an original problem of quantile hedging.

The paper is organized as follows. In section 2 we precisely
formulate the problem. It turns out that the solution on complete
markets has a clear economic interpretation. It is presented in
section 3. Sections 4 and 5 provide examples of Black-Scholes
model and the CRR model. For B-S model explicit solution is found
while for the CRR model existence is clear, but solutions are
found for some particular cases. In section 6 result for
incomplete markets is proved and presented in a one step trinomial
model.

\section{Problem formulation}
We consider financial markets with either discrete or continuous
time and with finite horizon $T$. Let $S_t$ be a $d$ dimensional
semimartingale describing evolution of stocks' prices on the
filtered probability space
$(\Omega,\mathcal{F},(\mathcal{F}_t),P)$. $X_t^{x,\pi}$ is a
wealth process connected with a pair $(x,\pi)$, where $\pi$ is a
predictable process describing self-financing strategy and $x$ is
an initial endowment. Thus the wealth process is defined by:
$X_0^{x,\pi}=x$, $X_t^{x,\pi}=\pi_t \cdot S_t$ and the
self-financing condition means that
\begin{align*}
\pi_t\cdot S_t&=\pi_{t+1}\cdot S_t &&\text{in case of discrete
time
model}\\
dX_t^{x,\pi}&=\pi_tdS_t, \quad \pi\in L(S) &&\text{in case of
continuous time model;} \ L(S) \text{is the set of predictable}\\
& && \text{processes integrable w.r. to S}.
\end{align*}
For simplicity assume that the interest rate is equal to zero and
that the set of all martingale measures $\mathcal{Q}$, so that
measures $Q$ that $S_t$ is a martingale with respect to $Q$ and
$Q\sim P$, is not empty. Among all self-financing strategies we
distinguish set $\mathcal{A}$ of all admissible strategies which
satisfy two additional conditions: $X_t^{x,\pi}\geq 0$ for all $t$
and $X_t^{x,\pi}$ is a supermartingale with respect to each
$Q\in\mathcal{Q}$. If $X_t^{x,\pi}\geq 0$ then the second
requirement is automatically satisfied for $S$ being a continuous
semimartingale, since then the wealth process is a $Q$-local
martingale bounded from below, so by Fatou's lemma it is a
supermartingale. In discrete time $X_t^{x,\pi}$ is even a
martingale, see $[5]$ Th. 2. Let $H$ be a nonnegative ,
$\mathcal{F}_T$ measurable random variable, called contingent
claim, which satisfies condition $H\in L^1(Q)$ for each
$Q\in\mathcal{Q}$. Its price at time $0$ is given by
$v_0=\sup_{Q\in\mathcal{Q}}\mathbf{E}^Q[H]$. This means that there
exists a strategy $\tilde{\pi}\in\mathcal{A}$ such that
$X_T^{v_0,\tilde{\pi}}\geq H$. Such $\tilde{\pi}$ is called a
hedging strategy. Now assume that we are given an initial capital
$0\leq x_0<v_0$. The question arises, what is an optimal strategy
for such endowment? As an optimality criterion we admit minimizing
probability of a large loss. Let $u:[0,\infty)\longrightarrow
[0,\infty)$ be a strictly increasing, continuous function such
that $u(0)=0$. Such function will be called a loss function. Let
$\alpha\geq 0$ be a level of acceptable losses. We are searching
for a pair $(x,\pi)$ such that
\begin{gather*}
P[u((H-X_T^{x,\pi})^{+})\leq\alpha]\longrightarrow
\max_{\pi\in\mathcal{A}},\\
x\leq x_0.
\end{gather*}
If there exists a solution $(x,\pi)$ of the problem above, then it
will be called optimal.

\section{Complete models}
Let $\mathcal{Q}=\{Q\}$, so the martingale measure is unique.
Recall that in this case each nonnegative $Q$~-~integrable
contingent claim $X$ can be replicated. It means that there exists
$\tilde{\pi}$ such that $X_T^{v_0,\tilde{\pi}}=X$, where
$v_0=\mathbf{E}^Q[X]$. In complete case the solution of our
problem has a clear economic interpretation. Let us start with the
basic theorem describing the solution.
\begin{tw}
If there exists $\tilde{X}\in L^0_+$ which is a solution of the
problem
\begin{gather*}
P[u((H-X)^+)\leq \alpha]\longrightarrow \max\\
\mathbf{E}^Q[X]\leq x_0
\end{gather*}
then the replicating strategy for $\tilde{X}$ is optimal.
\end{tw}
{\bf Proof :}\quad Recall that for $(x,\pi), \pi\in\mathcal{A}$
the wealth process  $X_t^{x,\pi}$ is a supermartingale with
respect to $Q$. Thus we have $\mathbf{E}^Q[X_T^{x,\pi}]\leq x\leq
x_0$ and $P[u((H-X_T^{x,\pi})^+)\leq
\alpha]\leq P[u((H-\tilde{X})^+)\leq \alpha]$.$\hfill{\square}$\\
\\
The main difficulty in this theorem is that we do not have an
existence result for $\tilde{X}$ and any method of constructing
which could be used for practical applications. However, we show
that the problem can be reduced to a simpler one by considering a
narrower class of random variables than $L_0^+$ and for this class
in some situations the problem can be explicitly solved. This is
an idea of considering strategies of class $\mathcal{S}$ which we
explain below.
\\ \\
{\bf Economic motivation for introducing strategies of class $\mathcal{S}$}\\
For $(x,\pi),\pi\in \mathcal{A}$ consider two sets:
$A=\{\omega\in\Omega : u((H-X_T^{x,\pi})^+)\leq\alpha\}$ and its
compliment $A^c$. Basing on $(x,\pi)$ let us build a modified
strategy $(\tilde{x},\tilde{\pi})$ in the following way. On $A$
investor's loss is smaller than $\alpha$. However, from our point
of view it can be as large as possible, but not larger than
$\alpha$. Therefore let $(\tilde{x},\tilde{\pi})$ be such that on
$A$ holds $u((H-X_T^{\tilde{x},\tilde{\pi}})^+)=\alpha$. On $A^c$
investor did not manage to hedge large loss, so the portfolio
value can be as well equal to $0$. Such $(\tilde{x},\tilde{\pi})$
we will regard as a strategy of class $\mathcal{S}$. What is an
advantage of such modification ? It turns out that
$\tilde{\pi}\in\mathcal{A}$ and the following inequalities hold:
\begin{gather*}
P[u((H-X_T^{\tilde{x},\tilde{\pi}})^{+})\leq\alpha]=P[u((H-X_T^{x,\pi})^{+})\leq\alpha],\qquad
\tilde{x}\leq x.
\end{gather*}
This fact is a motivation for searching the solution of the
problem only among strategies of class $\mathcal{S}$. Below we
present this idea in a more precise way.
\begin{dfn}
Random variable $X\in L^0_{+}$ is of class $\mathcal{S}$ if there
exists $A\in\mathcal{F}$ containing $\{u(H)\leq\alpha\}$ such that
\begin{enumerate}
\item on $A$ we have
\begin{enumerate}
\item if $u(H)\leq\alpha$ then $X=0$

\item if $u(H)>\alpha$ then $u(H-X)=\alpha$
\end{enumerate}
\item on $A^c$ we have $X=0$.
\end{enumerate}
\end{dfn}
Notice that on the set $A$ we have $X=0$ if $H\leq u^{-1}(\alpha)$
and $X=H-u^{-1}(\alpha)$ if $H>u^{-1}(\alpha)$. Thus on $A$ we
have $X=(H-u^{-1}(\alpha))^+$. Since $X=0$ on $A^c$ we obtain that
$X=\mathbf{1}_A(H-u^{-1}(\alpha))^+$. In other words
$X\in\mathcal{S}$ if it is of the form
$X=\mathbf{1}_A(H-u^{-1}(\alpha))^+$ for some $A\in\mathcal{F}$
such that $A\supseteq\{u(H)\leq\alpha\}$.

\begin{lem}
For each $X\in L^0_{+}$ such that $\mathbf{E}^Q[X]\leq x_0$ there
exists a random variable $Z\in\mathcal{S}$ such that
$\mathbf{E}^Q[Z]\leq\mathbf{E}^Q[X]$ and
\begin{gather*}
P[u((H-X)^{+})\leq\alpha]=P[u((H-Z)^{+})\leq\alpha].
\end{gather*}
\end{lem}
{\bf Proof : } Let us define $A:=\{\omega: u(H-X)^+\leq \alpha\}$.
Then $Z:=\mathbf{1}_A(H-u^{-1}(\alpha))^+\in \mathcal{S}$ and we
have
\begin{align*}
P[u((H&-Z)^+)\leq
\alpha]=P[u((H-\mathbf{1}_A(H-u^{-1}(\alpha))^+)^+)\leq \alpha]\\
&=P[\omega\in A:u((H-(H-u^{-1}(\alpha))^+)^+)\leq
\alpha]+P[\omega\in
A^c:u(H)\leq \alpha]\\
&=P[\omega\in A\cap\{u(H)\leq\alpha\}:u(H)\leq \alpha]+P[\omega\in
A\cap\{u(H)>\alpha\} :u(u^{-1}(\alpha)))\leq \alpha]\\
&=P(A).
\end{align*}
On the set $A^c$ holds $Z=0\leq X$. On $A$ if $u(H)\leq\alpha$
then $Z=0\leq X$ and if $u(H)>\alpha$ then $Z=H-u^{-1}(\alpha)\leq
X$. Thus we have $Z\leq X$ and $E^{Q}[Z]\leq E^{Q}[X]$.
$\hfill{\square}$\\

\begin{rem}
The above calculations show that for any
$X=\mathbf{1}_B(H-u^{-1}(\alpha))^+\in\mathcal{S}$ holds
\begin{gather*}
P(u(H-X)^+\leq\alpha)=P(B).
\end{gather*}
\end{rem}
Using lemma $3.3$ and remark $3.4$ we can reformulate theorem
$3.1$ in the following form.

\begin{tw}
If there exists set $\tilde{A}\supseteq \{u(H)\leq\alpha \}$ which
is a solution of the problem :
\begin{gather}
P(A)\longrightarrow \max\\
E^{Q}[\mathbf{1}_A(H-u^{-1}(\alpha))^+]\leq x_0
\end{gather}
then the replicating strategy for
$\mathbf{1}_{\tilde{A}}(H-u^{-1}(\alpha))^+$ is optimal.
\end{tw}
{\bf Proof :} Indeed, by lemma $3.3$ the problem
\begin{gather*}
P[u(H-X)^{+}\leq\alpha]\longrightarrow\max,\quad
\mathbf{E}^Q[X]\leq x_0,\quad X\in L^0_+
\end{gather*}
can be replaced by
\begin{gather*}
P[u(H-X)^{+}\leq\alpha]\longrightarrow\max,\quad
\mathbf{E}^Q[X]\leq x_0,\quad X\in \mathcal{S}.
\end{gather*}
However, by remark $3.4$ we know that for
$X=\mathbf{1}_A(H-u^{-1}(\alpha))^{+}\in\mathcal{S}$ we have\\
$P[u(H-X)^{+}\leq\alpha]=P(A)$ and the required formulation is
obtained. $\hfill{\square}$\\

\begin{rem}
Let us consider the optimizing problem from theorem $3.5$ given by
$3.5.1$ and $3.5.2$ but without the requirement that
$\tilde{A}\supseteq\{u(H)\leq\alpha\}$. Notice that if
$P(u(H)\leq\alpha)>0$ then the solution $\tilde{A}$ must contain
$\{u(H)\leq\alpha\}$. Suppose the contrary and define
$\tilde{\tilde{A}}:=\tilde{A}\cup\{u(H)\leq\alpha\}$. Then
$\mathbf{E}^Q[\mathbf{1}_{\tilde{\tilde{A}}}(H-u^{-1}(\alpha))^{+}]=
\mathbf{E}^Q[\mathbf{1}_{\tilde{A}}(H-u^{-1}(\alpha))^{+}]\leq x_0
$ and $P(\tilde{\tilde{A}})>P(\tilde{A})$ what is a contradiction.
This shows that that the requirement
$\tilde{A}\supseteq\{u(H)\leq\alpha\}$ in the theorem 3.5 can be
dropped.
\end{rem}
\noindent
In some particular cases the existence and construction
of the set $\tilde{A}$ can be solved by using Neyman-Pearson
lemma. To this end let us introduce a measure $\bar{Q}$ which is
absolutely continuous with respect to $Q$ by:
\begin{gather*}
\frac{d\bar{Q}}{dQ}=\frac{(H-u^{-1}(\alpha))^+}{\mathbf{E}^Q[(H-u^{-1}(\alpha))^+]}.
\end{gather*}
Then set $\tilde{A}$ solves the following problem
\begin{gather*}
P(A)\longrightarrow \max\\
\bar{Q}(A)\leq \frac{x_0}{\mathbf{E}^Q[(H-u^{-1}(\alpha))^+]}.
\end{gather*}
To make the paper self-contained  we present a part of the
Neyman-Pearson lemma. Let $P_1$ and $P_2$ be two probability
measures such that there exists density $\frac{dP_1}{dP_2}$.
\begin{lem}
If there exists constant $\beta$ such that
$P_2\{\frac{dP_1}{dP_2}\geq\beta\}=\gamma$ then
$P_1\{\frac{dP_1}{dP_2}\geq\beta\}\geq P_1(B)$ for any set $B$
satisfying $P_2(B)\leq\gamma$.
\end{lem}
{\bf Proof :} Let $B$ be a set satisfying $P_2(B)\leq\gamma$ and
denote $\tilde{B}:=\{\frac{dP_1}{dP_2}\geq\beta\}$. Then we have
\begin{align*}
P_1(\tilde{B})-P(B)&=\int_{\Omega}(\mathbf{1}_{\tilde{B}}-\mathbf{1}_B)dP_1=
\int_{\frac{dP_1}{dP_2}\geq\beta}(\mathbf{1}_{\tilde{B}}-\mathbf{1}_B)dP_1+
\int_{\frac{dP_1}{dP_2}<\beta}(\mathbf{1}_{\tilde{B}}-\mathbf{1}_B)dP_1\\[2ex]
&\geq
\int_{\frac{dP_1}{dP_2}\geq\beta}(\mathbf{1}_{\tilde{B}}-\mathbf{1}_B)\beta
dP_2-
\int_{\frac{dP_1}{dP_2}<\beta}\mathbf{1}_{B} \beta dP_2\\[2ex]
& = \beta\left(\int_{\tilde{B}} dP_2- \int_{B} dP_2\right)
=\beta(\gamma-P_2(B))\geq 0.
\end{align*}
$\hfill{\square}$\\
This lemma is useful for the Black-Scholes model since there the
condition
$\bar{Q}\{\frac{dP}{d\bar{Q}}\geq\beta\}=\frac{x_0}{\mathbf{E}^Q[(H-u^{-1}(\alpha))^+]}$
is satisfied. However, in case of discrete $\Omega$ this condition
no longer holds. This will be shown in the example of the $CRR$
model.

\section{Black-Scholes model}
Here we follow an example presented in $[3]$. The stock price
$S_t$ is given by
\begin{gather*}
dS_t=S_t(\mu dt + \sigma dW_t),\quad S_0=s,
\end{gather*}
where $\mu$ and $\sigma>0$ are constants and $W_t$ is a standard
Brownian motion. For this model
\begin{gather*}
S_t=se^{(\mu-\frac{1}{2}\sigma^2)t+\sigma W_t}.
\end{gather*}
and the unique martingale measure $Q$ is given by
\begin{gather*}
\frac{dQ}{dP}=e^{-\frac{\mu}{\sigma}W_T-\frac{1}{2}(\frac{\mu}{\sigma})^2T}.
\end{gather*}
Moreover, the process $W_t^\ast=W_t+\frac{\mu}{\sigma}t$ is a
Brownian motion with respect to $Q$. Notice that the density of
the martingale measure can be expressed in term of $S_T$, namely
\begin{gather*}
\frac{dQ}{dP}=cS_T^{-\frac{\mu}{\sigma^2}}, \quad \text{where} \ c
\text{ \ is some constant}.
\end{gather*}
We study a risk minimizing problem for a European call option with
strike $K$. Recall that the problem is reduced to constructing set
$\tilde{A}$ being a solution of
\begin{gather*}
P(A)\longrightarrow\max\\[2ex]
\bar{Q}(A)\leq\frac{x_0}{\mathbf{E}^Q[(S_T-K-u^{-1}(\alpha))^{+}]},
\end{gather*}
where measure $\bar{Q}$ is as in the previous section :
\begin{gather*}
\frac{d\bar{Q}}{dQ}=\frac{(S_T-K-u^{-1}(\alpha))^+}{\mathbf{E}^Q[(S_T-K-u^{-1}(\alpha))^+]}.
\end{gather*}
Notice, that the superscript $"+"$ above can be dropped since for
any $a,b,c\geq 0$ holds $((a-b)^+-c)^+=(a-b-c)^+.$ According to
Neyman-Pearson lemma we are searching for the set $\tilde{A}$ of
the form:
\begin{gather*}
\bigg\{\frac{dP}{d\bar{Q}}\geq c_1\bigg\}
=\bigg\{\frac{dP}{dQ}\geq c_2
(S_T-K-u^{-1}(\alpha))^+\bigg\}=\bigg\{S^{\frac{\mu}{\sigma^2}}_T\geq
c\cdot c_2(S_T-K-u^{-1}(\alpha))^+\bigg\},
\end{gather*}
where $c_1,c_2$ are nonnegative constants such that
\begin{gather}
E^Q[\mathbf{1}_{\tilde{A}}(S_T-K-u^{-1}(\alpha))^+]=x_0.
\end{gather}
Let us
consider two cases.
\\
$1)$ $\mu\leq\sigma^2$\\
Then the function $x\longrightarrow x^{\frac{\mu}{\sigma^2}}$ is
concave and has $0$ in $0$ and thus the solution is given by

$\tilde{A}=\{S_T\leq c_3\}=\{W^\ast_T\leq c_4\}$, where $c_3$ and
$c_4$ s.t. $c_3=se^{\sigma c_4-\frac{1}{2}\sigma^2 T}$ are
constant numbers satisfying $3.0.3$. The optimal strategy is a
strategy which replicates the following contingent claim:
\begin{align*}
\mathbf{1}_{\tilde{A}}(S_T-K-u^{-1}(\alpha))^+&=\mathbf{1}_{\{S_T\leq
c_3\}}(S_T-K-u^{-1}(\alpha))^{+}\\[2ex]
&=(S_T-K-u^{-1}(\alpha))^{+}-(S_T-c_3)^{+}-(c_3-K-u^{-1}(\alpha))\mathbf{1}_{\{S_T>c_3\}}
\end{align*}
and the corresponding probability is equal
\begin{gather*}
P(\tilde{A})=P(W_T^{\ast}\leq
c_4)=\Phi\left(\frac{c_4-\frac{\mu}{\sigma}T}{\sqrt{T}}\right).
\end{gather*}
For calculating constants $c_3$ and $c_4$ from $4.0.3$ we use
formula for pricing European call option.
\begin{align*}
&\mathbf{E}^Q\Big[(S_T-K-u^{-1}(\alpha))^{+}-(S_T-c_3)^{+}-(c_3-K-u^{-1}(\alpha))\mathbf{1}_{\{S_T>c_3\}}\Big]=\\[2ex]
&s\Phi(\bar{d}_{+})-(K+u^{-1}(\alpha))\Phi(\bar{d}_{-})-s\Phi\bigg(\frac{-c_4+\sigma
T}{\sqrt{T}}\bigg)+c_3\Phi\bigg(-\frac{c_4}{\sqrt{T}}\bigg)-\\[2ex]
&(c_3-K-u^{-1}(\alpha))Q\{W^{\ast}_T>c_4\}=\\[2ex]
&s\Phi(\bar{d}_{+})-(K+u^{-1}(\alpha))\Phi(\bar{d}_{-})-s\Phi\bigg(\frac{-c_4+\sigma
T}{\sqrt{T}}\bigg)+(K+u^{-1}(\alpha))\Phi\bigg(-\frac{c_4}{\sqrt{T}}\bigg)=x_0,
\end{align*}
where $\bar{d}_{{+}\over{}}=-\frac{1}{\sigma\sqrt{T}}
\ln\Big(\frac{K+u^{-1}(\alpha)}{s}\Big){{+}\over{}}\frac{1}{2}\sigma\sqrt{T}$
and $\Phi$ stands for the distribution function of the $N(0,1)$
distribution. \\ \noindent
$2)$ $\mu>\sigma^2$ \\
In this case the function $x\longrightarrow
x^{\frac{\mu}{\sigma^2}}$ is convex and therefore our solution is
of the form
\begin{gather*}
\tilde{A}=\{S_T<c_5\}\cup\{S_T>c_6\}=\{W^\ast_T<c_7\}\cup\{W^\ast_T>c_8\}
\end{gather*}
where $c_5<c_6$ are two solutions of the equation
$x^{\frac{\mu}{\sigma^2}}=\bar{c}(x-K-u^{-1}(\alpha))^{+}$, where
$\bar{c}$ is a constant number s.t. $4.0.3$ holds. Constants
$c_7,c_8$ are given by $c_5=se^{\sigma c_7-\frac{1}{2}\sigma^2
T},c_6=se^{\sigma c_8-\frac{1}{2}\sigma^2 T}$. The optimal
strategy is a strategy which replicates the following contingent
claim:
\begin{align*}
&\mathbf{1}_{\tilde{A}}(S_T-K-u^{-1}(\alpha))^+=\\[2ex]
&(S_T-K-u^{-1}(\alpha))^{+}-(S_T-c_5)^{+}-(c_5-K-u^{-1}(\alpha))\mathbf{1}_{\{S_T>c_5\}}+
(S_T-c_6)^{+}+\\[2ex]
&(c_6-K-u^{-1}(\alpha))\mathbf{1}_{\{S_T>c_6\}}
\end{align*}
and the corresponding probability is equal
\begin{gather*}
P(\tilde{A})=P(W_T^{\ast}<c_7)+P(W_T^{\ast}>c_8)=
\Phi\bigg(\frac{c_7-\frac{\mu}{\sigma}T}{\sqrt{T}}\bigg)+\Phi\bigg(-\frac{c_8-\frac{\mu}{\sigma}T}{\sqrt{T}}\bigg).
\end{gather*}
Now we need to determine all necessary constants. Using the same
methods as in the previous case we obtain
\begin{align}
\nonumber&\mathbf{E}^Q[\mathbf{1}_{\tilde{A}}(S_T-K-u^{-1}(\alpha))^+]=s\Phi(\bar{d}_{+})-(K+u^{-1}(\alpha))\Phi(\bar{d}_{-})
-\\[2ex]
&s\Phi\bigg(-\frac{c_7}{\sqrt{T}}+\sigma\sqrt{T}\bigg)+s\Phi\bigg(-\frac{c_8}{\sqrt{T}}+\sigma\sqrt{T}\bigg)+
\Big(K+u^{-1}(\alpha)\Big)\bigg(\Phi\bigg(-\frac{c_7}{\sqrt{T}}\bigg)-\Phi\bigg(-\frac{c_8}{\sqrt{T}}\bigg)\bigg)=x_0
\end{align}
Summarizing, constants are determined by $4.0.4$ and by the fact
that $c_5,c_6$ are solutions of the equation
$x^{\frac{\mu}{\sigma^2}}=\bar{c}(x-K-u^{-1}(\alpha))^{+}$, where
$\bar{c}$ is a positive constant.

\section {CRR model}
Let $(S_n)_{n=0,1,2,...,N}$ be a stock price given by
\begin{gather*}
S_{n+1}=S_n(1+\rho_n),\quad S_0=S,
\end{gather*}
where $(\rho_n)$ is a sequence of independent random variables
such that $p:=P(\rho_n=u)=1-P(\rho_n=d)$, where $u>d, u>0, d<0 $.
This means that at any time the price process $S_n$ can increase
to the value $S_n(1+u)$ or decrease to $S_n(1+d)$. We assume that
$p\in(0,1)$. It is known, that the unique martingale measure for
this model is given by $p^\ast:=\frac{-d}{u-d}$.\\
Let us study the risk minimizing problem for the call option with
strike $K$. Let us denote
$(S_N-\bar{K})^+:=(S_N-K-u^{-1}(\alpha))^+$ and consider two
measures: the objective one $P$
\begin{gather*}
P(\omega_k)=p^k(1-p)^{N-k}\\
\end{gather*}
and the measure $\bar{Q}$ (which is not necessarily a probability
measure) given by
\begin{gather*}
\bar{Q}(\omega_k):=(S(1+u)^k(1+d)^{N-k}-\bar{K})^+{p^\ast}^k(1-p^\ast)^{N-k}.
\end{gather*}
Here $\omega_k$ means an elementary event for which the number of
jumps upwards is equal to $k$. Our aim is to find set $\tilde{A}$
which solves:
\begin{gather*}
P(A)\longrightarrow\max\\
\bar{Q}(A)\leq x_0.
\end{gather*}
For the $CRR$ model existence of the required set $\tilde{A}$ is
clear since $\Omega$ is finite. However we want to find it
explicitly. Unfortunately, the Neyman-Pearson lemma for the
measures $P$ and $\bar{Q}$ can not be applied here since $\Omega$
is discrete and the condition
\begin{gather*}
\bar{Q}\bigg\{\frac{dP}{dQ}\geq
a(H-u^{-1}(\alpha))^+\bigg\}=\frac{x_0}{E[(H-u^{-1}(\alpha))^+]}
 \ \text{for some} \ a>0
\end{gather*}
is very rarely satisfied. The first way of constructing
$\tilde{A}$, which seems to be natural, is to find a constant
$\bar{a}$ such that
\begin{gather*}
\bar{a}=\inf\bigg\{a: \bar{Q} \Big\{\frac{dP}{dQ}\geq
a(H-u^{-1}(\alpha))^+\Big\}\leq
\frac{x_0}{\mathbf{E}^Q[(H-u^{-1}(\alpha))^+]}\bigg\}.
\end{gather*}
and then expect that
\begin{gather*}
\bar{A}=\bigg\{\frac{dP}{dQ}\geq
\bar{a}(H-u^{-1}(\alpha))^+\bigg\}
\end{gather*}
is a solution. Unfortunately, this is not a right construction as
shown in the example below.
\\
{\bf Example} \\
Let $\Omega=\{\omega_1,\omega_2,\omega_3\}$ and $P$ and $Q$ are
two measures given by
$p_1=\frac{7}{15},p_2=\frac{4}{15},p_3=\frac{4}{15}$ and
$q_1=\frac{4}{10},q_2=\frac{3}{10},q_3=\frac{3}{10}$. We want to
maximize $P(A)$ subject to the condition $Q(A)\leq
x_0=\frac{6}{10}$. We have
$\frac{p_1}{q_1}=\frac{63}{54},\frac{p_2}{q_2}=\frac{48}{54},\frac{p_3}{q_3}=\frac{48}{54}$
and the above construction gives $\tilde{A}=\{\omega_1\}$. However
$Q(\{\omega_2,\omega_3\})=\frac{6}{10}$ and
$P(\{\omega_2,\omega_3\})=\frac{8}{15}>\frac{7}{15}=P(\omega_1)$.
\\
\\
Below we present a lemma which provides construction o $\tilde{A}$
when measures satisfy some particular condition. It turns out that
this condition is satisfied by a significant number of cases in
the hedging problem of call option.
\begin{lem}
Let $\Omega=\{\omega_1,\omega_2,...,\omega_n\}$ and measures $P$
and $Q$ (not necessary probabilistic) satisfy the following
conditions:
\begin{gather*}
p_1\geq p_2\geq p_3\geq...\geq p_n >0< q_1\leq q_2\leq
q_3\leq...\leq q_n
\end{gather*}
and $\gamma$ be a fixed constant. Let
$\tilde{A}=\{\omega_1,\omega_2,...,\omega_k \}$, where the number
$k$ is such that $Q(\omega_1,\omega_2,...,\omega_k)\leq \gamma$
and $Q(\omega_1,\omega_2,...,\omega_k,\omega_{k+1})> \gamma$. Then
$P(\tilde{A})\geq P(A)$ for any set $A$ satisfying $Q(A)\leq
\gamma$.
\end{lem}
{\bf Proof :}\\
Let $B\subseteq\Omega$ s.t. $Q(B)\leq\gamma$.\\
$1)$ First assume that $\tilde{A}\cap B=\emptyset$. Then $\mid
B\mid\leq k$ and we have $P(\tilde{\omega})\geq P(\omega)$ for
each $\tilde{\omega}\in\tilde{A}$ and $\omega\in B$. As a
consequence
\begin{gather*}
P(\tilde{A})=\sum_{\omega\in \tilde{A}}P(\omega)\geq
\sum_{\omega\in B}P(\omega)=P(B).
\end{gather*}
$2)$ If $\tilde{A}\cap B \neq\emptyset$ then by $(1)$
$\tilde{A}\setminus\{\tilde{A}\cap B\}$ is a solution of
\begin{gather*}
P(A)\longrightarrow\max\\[2ex]
Q(A)\leq \gamma-Q(\{\tilde{A}\cap B\})
\end{gather*}
and so $P(\tilde{A}\setminus\{\tilde{A}\cap B\})\geq
P(B\setminus\{\tilde{A}\cap B\})$. As a consequence
$P(\tilde{A})\geq P(B)$. $\hfill\square$\\
\\
\noindent Since $P(\omega_k)$ increases with $k$ if
$p>\frac{1}{2}$ and decreases if $p<\frac{1}{2}$, the only point
to apply the lemma is to state the monotonicity of the measure
$\bar{Q}$. In fact we are interested in monotonicity of $\bar{Q}$
only on the set where it is strictly positive. Let us denote
\begin{align*}
a_k&:=\bar{Q}(\omega_k)=(S(1+u)^k(1+d)^{N-k}-\bar{K})^+{p^\ast}^k(1-p^\ast)^{N-k},\\[2ex]
b_k&:=\frac{(S(1+u)^{k+1}(1+d)^{N-k-1}-\bar{K})^+}{(S(1+u)^{k}(1+d)^{N-k}-\bar{K})^+},\\[2ex]
q&:=\frac{1+u}{1+d} \ ,
\end{align*}
where the sequence $b_k$ is well defined under convention that
$\frac{a}{0}=\infty$ for $a\geq 0$. Then $\bar{Q}(\omega_k)$ is
increasing if $\frac{a_{k+1}}{a_k}\geq 1$ for each
$k=0,1,...,N-1$. This condition is equivalent to that $b_k\geq
\frac{1-p^\ast}{p^\ast}$ for each $k=0,1,...,N-1$. But now note
that the sequence $b_k$ is decreasing. To see that one can
calculate that
\begin{gather*}
\frac{b_{k+1}}{b_k}\leq 1 \ \Longleftrightarrow \ (q-1)^2\geq 0 .
\end{gather*}
The last condition is always satisfied. Thus $\bar{Q}(\omega_k)$
is increasing if
\begin{gather*}
b_{N-1}=\frac{(S(1+u)^N-\bar{K})^+}{(S(1+u)^{N-1}(1+d)-\bar{K})^+}\geq\frac{1-p^\ast}{p^\ast}.
\end{gather*}
Note that this case includes the situations when
$p^\ast\geq\frac{1}{2}$. \\
By analogous arguments one can obtain condition under which
$\bar{Q}(\omega_k)$ is decreasing. This is the case when the
$b_{\bar{k}}\leq \frac{1-p^\ast}{p^\ast}$, where $\bar{k}$ is the
minimal $k$ for which $b_k\neq\infty$. Indeed, then we have
$b_k\leq\frac{1-p^\ast}{p^\ast}$ for all $k\geq\bar{k}$ what
implies that $a_{k+1}<a_k$ for $k\geq\bar{k}$. \\
Before summarizing the above consideration let us introduce the
following notation
\begin{gather*}
A_k:=\{\omega\in\Omega \text{ \ s.t. the number of jumps upwards
is equal to } k \}.
\end{gather*}
for the set containing all elements $\omega_k$. The following
lemma is a consequence of lemma $5.1$.
\begin{lem} \quad\\
$1)$(P increasing, $\bar{Q}$ decreasing) \\
Let $\bar{k}=\min\{k: b_k\neq\infty\}$. If $p\geq\frac{1}{2}$ and
$b_{\bar{k}}\leq \frac{1-p^\ast}{p^\ast}$ then
$\tilde{A}=A_{N}\cup A_{N-1}\cup...\cup A_{N-k}\cup B_{N-k-1}$,
where the number $k$ is s.t. $\bar{Q}(A_{N}\cup A_{N-1}\cup...\cup
A_{N-k})\leq x_0$ and $\bar{Q}(A_{N}\cup A_{N-1}\cup...\cup
A_{N-k-1})> x_0$ and the set $B_{N-k-1}$ contains maximal number
of any elements from the set $A_{N-k-1}$ such that
$\bar{Q}(B_{N-k-1})\leq x_0 - \bar{Q}(A_{N}\cup A_{N-1}\cup...\cup
A_{N-k})$.
\\
\noindent $2)$(P decreasing, $\bar{Q}$ increasing) \\
If $p\leq\frac{1}{2}$ and
$\frac{(S(1+u)^N-\bar{K})^+}{(S(1+u)^{N-1}(1+d)-\bar{K})^+}\geq
\frac{1-p^\ast}{p^\ast}$ (for example when
$p^\ast\geq\frac{1}{2}$)  then $\tilde{A}=A_{0}\cup
A_{1}\cup...\cup A_{k}\cup B_{k+1}$, where the number $k$ is s.t.
$\bar{Q}(A_{0}\cup A_{1}\cup...\cup A_{k})\leq x_0$ and
$\bar{Q}(A_{0}\cup A_{1}\cup...\cup A_{k+1})> x_0$ and the set
$B_{k+1}$ contains maximal number of any elements from the set
$A_{k+1}$ such that $\bar{Q}(B_{k+1})\leq x_0 - \bar{Q}(A_{0}\cup
A_{1}\cup...\cup A_{k})$.

\end{lem}

\noindent
{\bf Example}\\
As an application of lemma $5.2$ we study a risk minimizing
problem for a call option with strike $K=600$ in a $3$-period
model with parameters : $S_0=1000$, $u=0,1$, $d=-0,2$,
$p=\frac{1}{4}$. Price at time $0$ of the option is
$u_0=\mathbf{E}^Q[(S_3-600)^{+}]=398\frac{7}{27}$. Assume that we
have only $x_0=150$ and $\alpha=5$ is a level of acceptable losses
measured by $u(x)=\sqrt{x}$. We denote by $\omega^{abc}$, where
$a,b,c \in \{u,d\}$ elementary events with interpretation of
$a,b,c$ as a history of the price process. For example
$\omega^{udu}$ means the event where the price process moved up in
the first and the third period and moved down in the second one.
Since we can not hedge the original contingent claim
$H=(S_3-600)^{+}$:
\begin{align*}
&H(\omega^{uuu})=731,\quad
H(\omega^{uud})=H(\omega^{udu})=H(\omega^{duu})=368,\\[2ex]
&H(\omega^{udd})=H(\omega^{dud})=H(\omega^{ddu})=104,\quad
H(\omega^{ddd})=0,
\end{align*}
we have to hedge $\tilde{H}=\mathbf{1}_{\tilde{A}}(S_3-625)^{+}$.
Since $p=\frac{1}{4}$ and $p^{\ast}=\frac{2}{3}$, we can apply
lemma $5.2(2)$ for construction of $\tilde{A}$. Below we present
three possible right candidates for $\tilde{H}$.
\begin{align*}
&\tilde{H}(\omega^{uuu})=0,\quad \tilde{H}(\omega^{ddd})=0,\quad
\tilde{H}(\omega^{ddu})=\tilde{H}(\omega^{dud})=\tilde{H}(\omega^{udd})=79\\[2ex]
&\quad\text{and} \
\Big\{\tilde{H}(\omega^{uud})=\tilde{H}(\omega^{udu})=343,\quad
\tilde{H}(\omega^{ddu})=0\Big\}\\[2ex]
&\quad\text{or} \ \Big\{\tilde{H}(\omega^{uud})=0,\quad
\tilde{H}(\omega^{udu})=\tilde{H}(\omega^{ddu})=343\Big\}\\[2ex]
&\quad\text{or} \ \Big\{\tilde{H}(\omega^{uud})=343,\quad
\tilde{H}(\omega^{udu})=0,\quad\tilde{H}(\omega^{ddu})=343\Big\}
\end{align*}
Moreover,
$P(\tilde{A})=\left(\frac{3}{4}\right)^3+\frac{1}{4}\cdot\left(\frac{3}{4}\right)^2\cdot
3+\left(\frac{1}{4}\right)^2\cdot\frac{3}{4}\cdot2=\frac{15}{16}$.

\section{Incomplete markets}
Now let us consider the case when the equivalent martingale
measure is not unique. This means that the market is incomplete
and not every contingent claim can be replicated. We preserve all
assumptions from previous section. Recall that the wealth process
$X_t^{x,\pi}$ is a supermartingale with respect to each martingale
measure $Q\in \mathcal{Q}$. In this case theorem which describes
optimal strategy is of the form:

\begin{tw}
Assume that there exists set $\tilde{A}$ which is a solution of
the problem:
\begin{align*}
P(A)\longrightarrow \max\\
\sup_{Q\in\mathcal{Q}}E^{Q}[\mathbf{1}_A(H-u^{-1}(\alpha))^+]\leq
x_0.
\end{align*}
Then the strategy which hedges the contingent claim
$\mathbf{1}_{\tilde{A}}(H-u^{-1}(\alpha))^+$ is optimal.
\end{tw}
{\bf Proof :}\\
Let us consider an arbitrary admissible strategy $(x,\pi)$, where
$x\leq x_0$. We will show that
$P(u(H-X_T^{x,\pi})^+\leq\alpha)\leq P(\tilde{A})$.\\
Notice, that for any $a,b,c\geq 0$ we have $(a-b)^+\leq c
\Longleftrightarrow b\geq(a-c)^+$ and thus
$u((H-X_T^{x,\pi})^+)\leq\alpha\Longleftrightarrow X_T^{x,\pi}\geq
(H-u^{-1}(\alpha))^+$. As a consequence for any $Q\in\mathcal{Q}$
we obtain
\begin{align*}
E^{Q}[\mathbf{1}_{\{u((H-X_T^{x,\pi})^+)\leq\alpha\}}(H-X_T^{x,\pi})^+]&\leq
E^{Q}[\mathbf{1}_{\{u((H-X_T^{x,\pi})^+)\leq\alpha\}}X_T^{x,\pi}]\\
&\leq E^{Q}[X_T^{x,\pi}]\leq x \leq x_0,
\end{align*}
where the last but one inequality follows from the fact that
$X_t^{x,\pi}$ is a $Q$ supermartingale. Taking supremum over all
martingale measures we have
\begin{gather*}
\sup_{Q\in\mathcal{Q}}E^{Q}[\mathbf{1}_{\{u((H-X_T^{x,\pi})^+)\leq\alpha\}}(H-u^{-1}(\alpha))^+]\leq
x_0.
\end{gather*}
 From the definition of the set $\tilde{A}$ we have
$P(u(H-X_T^{x,\pi})^+\leq\alpha)\leq P(\tilde{A})$. \\
Now let us consider the strategy $(\tilde{x},\tilde{\pi})$ which
hedges $\mathbf{1}_{\tilde{A}}(H-u^{-1}(\alpha))^+$. We have
\begin{gather*}
\{u(H-X_T^{\tilde{x},\tilde{\pi}})^+\leq\alpha\}=\{X_T^{\tilde{x},\tilde{\pi}}\geq(H-u^{-1}(\alpha))^+\}\supseteq
\{X_T^{\tilde{x},\tilde{\pi}}\geq\mathbf{1}_{\tilde{A}}(H-u^{-1}(\alpha))^+\}\supseteq
\tilde{A}
\end{gather*}
and so $P(u(H-X_T^{\tilde{x},\tilde{\pi}})^+\leq\alpha)\geq
P(\tilde{A})$. It follows that $(\tilde{x},\tilde{\pi})$ is
optimal and moreover we have
$P(u(H-X_T^{\tilde{x},\tilde{\pi}})^+\leq\alpha)=P(\tilde{A})$.\hfill{$\square$}
\\
\\
The main problem which needs to be investigated is the existence
of the set $\tilde{A}$. We are not in a position to prove a
general existence result for $\tilde{A}$ but we will show an
example of trinomial model where it can be explicitly found.
\\
\\
\noindent
{\bf Example - Trinomial model} \\
Let us consider a one-step model where the stock price is given by
\begin{gather*}
S_1=S(1+\xi), \ \text{where} \ P(\xi=a)=p_1, \ P(\xi=b)=p_2, \
P(\xi=c)=p_3\\
 a>b>c, \ p_1,p_2,p_3>0,  \ p_1+p_2+p_3=1
\end{gather*}
and where the interest rate is equal to $0$. Here $S$ is an
initial price and $S_1$ is a price at time $1$. To obtain the
arbitrage-free model we assume that $a>0$ and $c<0$. Contingent
claim is denoted by
$H=(H_1,H_2,H_3)=(H(\omega_1),H(\omega_2),H(\omega_3))$.
\\
First let us study the structure of the set of all martingale
measures $\mathcal{Q}$. Each $Q\in\mathcal{Q}$ is a triplet
$Q=(q_1,q_2,q_3)$ which is a solution of the system
\[
\begin{cases}
\quad q_1S_0(1+a)+q_2S_0(1+b)+q_3S_0(1+c)&=S_0 \\
\hfill q_1+q_2+q_3&=1 \\
\hfill q_1,q_2,q_3&>0 \ .
\end{cases}
\]
By direct computation we obtain that such triplet can be
parametrized by $q_1$. Precisely speaking each martingale measure
is of the form:
\begin{gather*}
Q=\left(q_1,\ \frac{c-a}{b-c}q_1+\frac{c}{c-b}, \
\frac{a-b}{b-c}q_1+\frac{b}{b-c}\right), \ \text{where} \
q_1\in(\underline{q},\bar{q}):=\left(0\vee\frac{b}{b-a}, \
\frac{c}{c-a}\right).
\end{gather*}
That means that each $Q\in\mathcal{Q}$ can be represented by
\begin{gather*}
Q=\alpha Q_1 + (1-\alpha) Q_2, \ \text{where} \ \alpha\in(0,1)
 \ \text{and}\\
Q_1=\left(\underline{q},\
\frac{c-a}{b-c}\underline{q}+\frac{c}{c-b}, \
\frac{a-b}{b-c}\underline{q}+\frac{b}{b-c}\right),\\
Q_2=\left(\bar{q},\ \frac{c-a}{b-c}\bar{q}+\frac{c}{c-b}, \
\frac{a-b}{b-c}\bar{q}+\frac{b}{b-c}\right).
\end{gather*}
Thus $\mathcal{Q}$ is a convex set with two vertexes $Q_1,Q_2$.
Now notice, that for any $A\in\mathcal{F}$ we have
\begin{gather*}
\sup_{Q\in\mathcal{Q}}E^Q[\mathbf{1}_{A}(H-u^{-1}(\alpha))^+]\leq
x_0
\ \text{if and only if} \\
 E^{Q_1}[\mathbf{1}_{A}(H-u^{-1}(\alpha))^+]\leq x_0 \quad \text{and}
 \quad
E^{Q_2}[\mathbf{1}_{A}(H-u^{-1}(\alpha))^+]\leq x_0,
\end{gather*}
 so the constraints for $\tilde{A}$ is reduced to two vertex measures.
As a consequence we are looking for a set $\tilde{A}$ which is a
solution of the problem
\begin{align*}
P(A)\longrightarrow \max
\end{align*}
\[
\begin{cases}
E^{Q_1}[\mathbf{1}_{A}(H-u^{-1}(\alpha))^+]\bar{Q}_1(A) \leq x_0\\
E^{Q_2}[\mathbf{1}_{A}(H-u^{-1}(\alpha))^+]\bar{Q}_2(A) \leq x_0 \
.
\end{cases}
\]
Now let us make concrete calculations for the case when $b>0$.
Then $Q_1=(0,\frac{c}{c-b},\frac{b}{b-c})$,
$Q_2=(\frac{c}{c-a},0,\frac{a}{a-c})$. Let
$\bar{H}:=(H-u^{-1}(\alpha))^+$,$\bar{H}_i:=(H_i-u^{-1}(\alpha))^+$.
Our problem is of the form:
\begin{align*}
\mathbf{1}_{\omega_1}(A)p_1+\mathbf{1}_{\omega_2}(A)p_2+\mathbf{1}_{\omega_3}(A)p_3\longrightarrow\max\\
\mathbf{1}_{\omega_2}(A)\frac{c}{c-b}\bar{H}_2+\mathbf{1}_{\omega_3}(A)\frac{b}{b-c}\bar{H}_3\leq x_0\\
\mathbf{1}_{\omega_1}(A)\frac{c}{c-a}\bar{H}_1+\mathbf{1}_{\omega_3}(A)\frac{a}{a-c}\bar{H}_3\leq x_0\\
\end{align*}
Since we do not have a general method of solving, we will check
all possibilities depending on $S,a,b,c,H,u,\alpha$. We will
denote by $L_1:=\frac{c}{c-b}\bar{H}_2+\frac{b}{b-c}\bar{H}_3$ and
by $L_2:=\frac{c}{c-a}\bar{H}_1+\frac{a}{a-c}\bar{H}_3$. We have
the following description of the set $\tilde{A}$.
\begin{enumerate}
%1
\item If $L_1\leq x_0$ and $L_2\leq x_0$ then
$\tilde{A}=\{\omega_1,\omega_3,\omega_3\}$.

%2
\item If $\min\{\frac{c}{c-b}\bar{H}_2,\frac{b}{b-c}\bar{H}_3\}>
x_0$ or
$\min\{\frac{c}{c-a}\bar{H}_1,\frac{a}{a-c}\bar{H}_3\}>x_0$ then
$\tilde{A}=\emptyset$.

%3
\item If $L_1\leq x_0$ and $L_2> x_0$ and
$\min\{\frac{c}{c-a}\bar{H}_1,\frac{a}{a-c}\bar{H}_3\}\leq x_0$
then if
\begin{enumerate}
\item $\max\{\frac{c}{c-a}\bar{H}_1,\frac{a}{a-c}\bar{H}_3\}> x_0$
and if
\begin{enumerate}
\item $\frac{c}{c-a}\bar{H}_1\geq\frac{a}{a-c}\bar{H}_3$ then
$\tilde{A}=\{\omega_2,\omega_3\}$, \item
$\frac{c}{c-a}\bar{H}_1<\frac{a}{a-c}\bar{H}_3$ then
$\tilde{A}=\{\omega_1,\omega_2\}$
\end{enumerate}
\item $\max\{\frac{c}{c-a}\bar{H}_1,\frac{a}{a-c}\bar{H}_3\}\leq
x_0$ and if
\begin{enumerate}
\item $p_1\geq p_3$ then $\tilde{A}=\{\omega_1,\omega_2\}$,

\item $p_1<p_3$ then $\tilde{A}=\{\omega_2,\omega_3\}$,
\end{enumerate}
\end{enumerate}

%4
\item If $L_1> x_0$ and $L_2\leq x_0$ and
$\min\{\frac{c}{c-b}\bar{H}_2,\frac{b}{b-c}\bar{H}_3\}\leq x_0$
then if
\begin{enumerate}
\item $\max\{\frac{c}{c-b}\bar{H}_2,\frac{b}{b-c}\bar{H}_3\}> x_0$
and if
\begin{enumerate}
\item $\frac{c}{c-b}\bar{H}_2\geq\frac{b}{b-c}\bar{H}_3$ then
$\tilde{A}=\{\omega_1,\omega_3\}$, \item
$\frac{c}{c-b}\bar{H}_2<\frac{b}{b-c}\bar{H}_3$ then
$\tilde{A}=\{\omega_1,\omega_2\}$
\end{enumerate}
\item $\max\{\frac{c}{c-b}\bar{H}_2,\frac{b}{b-c}\bar{H}_3\}\leq
x_0$ and if
\begin{enumerate}
\item $p_2\geq p_3$ then $\tilde{A}=\{\omega_1,\omega_2\}$,

\item $p_2<p_3$ then $\tilde{A}=\{\omega_1,\omega_3\}$,
\end{enumerate}
\end{enumerate}

%5
\item If $L_1> x_0$ and $L_2> x_0$ and
$\min\{\frac{c}{c-b}\bar{H}_2,\frac{b}{b-c}\bar{H}_3\}\leq x_0$
and $\min\{\frac{c}{c-a}\bar{H}_1,\frac{a}{a-c}\bar{H}_3\}\leq
x_0$ then if
\begin{enumerate}
\item $\max\{\frac{c}{c-b}\bar{H}_2,\frac{b}{b-c}\bar{H}_3\}> x_0$
and $\max\{\frac{c}{c-a}\bar{H}_1,\frac{a}{a-c}\bar{H}_3\}> x_0$
and if
\begin{enumerate}
\item $\frac{c}{c-b}\bar{H}_2\leq\frac{b}{b-c}\bar{H}_3$ and
$\frac{c}{c-a}\bar{H}_1\leq\frac{a}{a-c}\bar{H}_3$ then
$\tilde{A}=\{\omega_1,\omega_2\}$

\item$\frac{c}{c-b}\bar{H}_2\leq\frac{b}{b-c}\bar{H}_3$ and
$\frac{c}{c-a}\bar{H}_1>\frac{a}{a-c}\bar{H}_3$ then
$\tilde{A}=\{\omega_2\}$

\item$\frac{c}{c-b}\bar{H}_2>\frac{b}{b-c}\bar{H}_3$ and
$\frac{c}{c-a}\bar{H}_1\leq\frac{a}{a-c}\bar{H}_3$ then
$\tilde{A}=\{\omega_1\}$

\item $\frac{c}{c-b}\bar{H}_2>\frac{b}{b-c}\bar{H}_3$ and
$\frac{c}{c-a}\bar{H}_1>\frac{a}{a-c}\bar{H}_3$ then
$\tilde{A}=\{\omega_3\}$
\end{enumerate}

\item$\max\{\frac{c}{c-b}\bar{H}_2,\frac{b}{b-c}\bar{H}_3\}> x_0$
and $\max\{\frac{c}{c-a}\bar{H}_1,\frac{a}{a-c}\bar{H}_3\}\leq
x_0$ and if
\begin{enumerate}
\item $\frac{c}{c-b}\bar{H}_2\leq\frac{b}{b-c}\bar{H}_3$ then
$\tilde{A}=\{\omega_1,\omega_2\}$

\item $\frac{c}{c-b}\bar{H}_2>\frac{b}{b-c}\bar{H}_3$ and if
\begin{enumerate}
\item $p_3\geq p_1$ then $\tilde{A}=\{\omega_3\}$

\item $p_3<p_1$ then $\tilde{A}=\{\omega_1\}$
\end{enumerate}
\end{enumerate}

\item $\max\{\frac{c}{c-b}\bar{H}_2,\frac{b}{b-c}\bar{H}_3\}\leq
x_0$ and $\max\{\frac{c}{c-a}\bar{H}_1,\frac{a}{a-c}\bar{H}_3\}>
x_0$ and if
\begin{enumerate}
\item $\frac{c}{c-a}\bar{H}_1\leq\frac{a}{a-c}\bar{H}_3$ then
$\tilde{A}=\{\omega_1,\omega_2\}$

\item $\frac{c}{c-a}\bar{H}_1>\frac{a}{a-c}\bar{H}_3$ and if
\begin{enumerate}
\item $p_2\geq p_3$ then $\tilde{A}=\{\omega_2\}$

\item $p_2<p_3$ then $\tilde{A}=\{\omega_3\}$
\end{enumerate}
\end{enumerate}

\item $\max\{\frac{c}{c-b}\bar{H}_2,\frac{b}{b-c}\bar{H}_3\}\leq
x_0$ and
$\max\{\frac{c}{c-a}\bar{H}_1,\frac{a}{a-c}\bar{H}_3\}\leq x_0$
and if

\begin{enumerate}
\item $p_1+p_2\geq p_3$ then $\tilde{A}=\{\omega_1,\omega_2\}$

\item $p_1+p_2<p_3$ then $\tilde{A}=\{\omega_3\}$.

\end{enumerate}

\end{enumerate}

\end{enumerate}

% bibliografia
%\thispagestyle{empty}
%\newpage
%\quad
\vskip 40pt {\LARGE{\bf References}}
\begin{description}
\item{[1]}\quad%{MB}
       M. Baran, \, \emph{Quantile hedging on markets with proportional transaction costs},
       Applicationes Mathematicae \ (2003), \ 193-208,
\item{[2]}\quad%{CK}
       J. Cvitanić, I. Karatzas\, \emph{On dynamic measures of risk},
       Finance and Stochastics 3 \ (1999), \ 451-482,
\item{[3]}\quad %{FL1}
     H. F\"{o}llmer, P. Leukert \, \emph{Quantile Hedging},
     Finance and Stochastics 3 \ (1999), \ 251-273,
\item{[4]}\quad%{FL2}
     H. F\"{o}llmer, P. Leukert \, \emph{Efficient Hedging:
     Cost versus Shortfall Risk}, \ Finance and Stochastics 4 \ (2000),\
     117-146,
\item{[5]}\quad%{P}
      J. Jacod, A.N. Shiryaev \, \emph{Local martingales and
     the fundamental asset pricing theorems in the discrete-time case}, \,
     Finance and Stochastics 2 (1998), \, 259-273
\item{[6]}\quad%{P}
     H. Pham \, \emph{Dynamic $L^{p}$-hedging in discrete time under cone
     constraints}, SIAM J. Control Optim. 38 \ (2000), No.3 \ 665-682.
\end{description}

\end{document}